# A self-consistent hybrid model of kinetic striations in low-current Argon discharges

Vladimir I. Kolobov,[1,2] Juan Alonso Guzman[2] and Robert R. Arslanbekov[1]

[1]CFD Research Corporation, Huntsville, AL, USA

[2]University of Alabama in Huntsville, Huntsville, AL, USA

**Abstract**

A self-consistent hybrid model of standing and moving striations was developed for low-current DC discharges in noble gases. We introduced the concept of surface diffusion in phase space (**r**,u) (where u denotes the electron kinetic energy) described by a tensor diffusion in the nonlocal Fokker-Planck kinetic equation for electrons in the collisional plasma. Electrons diffuse along surfaces of constant total energy $\varepsilon = u - e\varphi(\boldsymbol{r})$ between energy jumps in inelastic collisions with atoms. Numerical solutions of the 1d1u kinetic equation for electrons were obtained by two methods and coupled to ion transport and Poisson solver. We studied the dynamics of striation formation in Townsend and glow discharges in Argon gas at low discharge currents using a two-level excitation-ionization model and a "full-chemistry" model, which includes stepwise and Penning ionization. Standing striations appeared in Townsend and glow discharges at low currents, and moving striations were obtained for the discharge currents exceeding a critical value. These waves originate at the anode and propagate towards the cathode. We have seen two types of moving striations with the 2-level and full-chemistry models, which resemble the *s* and *p* striations previously observed in the experiments. Simulations indicate that processes in the anode region could control moving striations in the positive column plasma. The developed model helps clarify the nature of standing and moving striations in DC discharges of noble gases at low discharge currents and low gas pressures.

## Introduction

Plasma stratification occurs in electric discharges of gases and gas mixtures over a wide range of pressures and discharge currents. Striations are ionization waves specific to collisional plasmas.[1,2] Studies of plasma stratification remain an area of active research testing our understating of electron kinetics and plasma chemistry.[3,4]

The most studied are striations in classical DC discharges of noble gases. Experimental observations obtained for over a century of studies produced a map of discharge states (*i/R*, *pR*) where *i* is the discharge current, *p* is the gas pressure, and *R* is the radius of the discharge tube. Scaling laws indicate that striations in vacuum discharges are similar to those observed in micro-discharges at high gas pressures. Several types of striations in noble gases have been identified. Striations in diffuse discharges at large *i/R* and low *pR* (near so-called Pupp boundary) are associated with a nonlinear dependence of the ionization rate on electron density, which appears due to Maxwellization of the electron energy distribution function caused by Coulomb interactions among charged particles. When volume recombination starts dominating over surface recombination (with increasing gas pressure), striations near the Pupp boundary gradually transit into two-dimensional waves propagating through radially constricted positive column plasma. At low $i/R$, three types of moving striations ($s$, $p,$ and $r$) have been observed under low $pR$. The peculiar feature of these striations is a specific electric potential drop over striation length related to the first excitation potential $\varepsilon_1$ of the noble gas atoms: one for $s$ striations, ½ for the $p$ striations, and 2/3 for the $r$ striations (Novak's law). Their nature originates from nonlocal kinetic effects in spatially periodic DC electric fields. With increasing $pR$ at low $i/R,$ the properties of the "kinetic" striations change as the nonlocal effects gradually vanish. Striations disappear when the Electron Energy Distribution Function (EEDF) becomes local, the radial plasma structure is controlled by volume recombination, and ionization non-linearities with respect to plasma density are absent.

Theoretical analysis of plasma stratification is usually performed by linearizing the system of governing equations with respect to small perturbations $\sim exp\, i\, (kx - \omega t)$ and determine the frequency $\omega$ and growth rate $\omega''$ as functions of the wave vector $k$ and discharge parameters (gas type and pressure, discharge current, tube radius, etc.).[5,6] The wavelength $\Lambda = 2\pi/k$ related to the maximum of growth rate $\omega''(k)$ corresponds to the observed striation wavelength in an infinitely long plasma. For $s$, $p,$ and $r$ striations, the maximum growth rate values $\omega''(k)$ are observed at the wavelengths corresponding to electric potential drop over striation length given by Novak's law.[7,8] The linear analysis does not consider the processes near electrodes and only applies to small-amplitude waves. The properties of nonlinear striations observed in experiments could be affected by the finite length of positive column plasma and the near-electrode processes, as discussed below.

Spatial relaxation of primary electrons emitted from the cathode and kinetic resonances in spatially periodic "striation-like" electric fields have been studied in numerous publications.[9] For a specific range of $E/N$, the ratio of the electric field and gas density, the spatial relaxation has the character of damped oscillations. Electron energy loss in elastic collisions and Coulomb interactions, excitation of several states of atoms, and generation of secondary electrons due to electron-impact ionization of atoms determine the spatial relaxation length. The production of secondary electrons in ionization events leads to a noticeable decrease in the relaxation length.[10]

Striations appear in PIC simulations of low-current discharges.[11] However, such simulations are not sufficient to fully clarify the nature of plasma stratification. Recent self-consistent fluid simulations of moving striations in DC discharges [12] confirmed the EEDF Maxwellization due to Coulomb collisions as the leading cause of Argon plasma stratification near the Pupp boundary. Similar nature of standing striations in RF discharges and moving striations in DC discharges of noble gases under similar conditions have been demonstrated.[13]

Attempts to develop self-consistent models of kinetic striations can be found in the literature. The authors of Ref. [14] proposed a hybrid model which includes the kinetic equation for electrons and the continuity equations for ions and excited atoms. An iterative procedure was used for self-consistent calculation of the electric field in a striated positive column under conditions when elastic collisions dominate in the electron energy balance, and Coulomb collisions are negligibly small. Arndt et al. [15] analyzed standing striations in a glow discharge by a coupled solution of spatially inhomogeneous kinetic equation for electrons, the hydrodynamic equations for ions and excited atoms, and the Poisson equation for the electric field. The authors of Ref. [16,17] attempted to describe the plasma stratification based on a coupled solution of the Boltzmann kinetic equation for electrons, a continuity equation for ions, and the Poisson equation for the self-consistent electric field. However, they neglected the source term in the ion balance equation by assuming that the ionization and recombination locally compensate each other. Furthermore, the ionization was treated as an inelastic collision in the kinetic equation that did not alter the number of particles. No surprise, the obtained solutions were strongly dependent upon the boundary condition on the cathode side of the positive column (PC). No self-consistent simulations of s, p, and r striations have been reported so far despite these efforts.

The present work aims at self-consistent modeling of the ionization waves in DC discharges of noble gases at low currents by a coupled solution of an electron kinetic equation, ion transport, and electrostatic field. The concept of surface diffusion in phase space is introduced and explored for solving the spatially inhomogeneous kinetic equation for electrons in a collisional plasma. The model is applied to studies of standing and moving striations in Argon at low currents. The modeling helps understand the formation of standing striations in the transitional region between the negative glow and PC and clarifies the differences between the moving and standing striations in DC discharges.

## Kinetic model for electrons

The electrons are modeled via a Fokker-Planck (FP) kinetic equation for the EEDF. This equation is derived through a two-term spherical harmonic expansion in velocity space:[18, 2]

$$\frac{\partial f_0}{\partial t} - \left(\nabla - \boldsymbol{E}\frac{\partial}{\partial u}\right) \cdot D_r \left(\nabla - \boldsymbol{E}\frac{\partial}{\partial u}\right) f_0 - \frac{1}{\sqrt{u}}\frac{\partial}{\partial u}\left(\sqrt{u}\Gamma_u\right) = C_0 \quad (1)$$

Here, $\boldsymbol{E}$ is the electric field, $u = mv^2/(2e)$ is the volt-equivalent of the electron kinetic energy, $D_r = v^2/(3\nu)$ is a diffusion coefficient in the $(r,u)$ phase space, and $\nu(v)$ is the transport collision frequency. The energy flux density $\Gamma_u$ describes dynamic friction and diffusion caused by collisions with small energy changes. The right-hand side of Eq. (1) contains inelastic collisions associated with significant energy changes per collision and the number of particles.

By neglecting the $\Gamma_u$ term, we can rewrite Eq. (1) as a diffusion equation in the 4-dimensional phase space $(r, u)$:

$$\frac{\partial f_0}{\partial t} + \nabla_4 \cdot (\boldsymbol{D}\nabla f) = C_0 \quad (2)$$

where **D** is the tensor diffusion coefficient

$$\boldsymbol{D} = D_r \begin{pmatrix} 1 & 0 & 0 & E_x \\ 0 & 1 & 0 & E_y \\ 0 & 0 & 1 & E_z \\ E_x & E_y & E_z & E^2 \end{pmatrix} \quad (3)$$

Both Dirichlet and Neumann type boundary conditions for Eq. (2) can be specified:[19]

$$\begin{aligned} \boldsymbol{D}\nabla f_0 &= \boldsymbol{g}_N \\ f_0 &= f_D \end{aligned} \qquad (4)$$

where $f_D$ is the prescribed value on the Dirichlet boundary, and $\boldsymbol{g}_N$ is a prescribed (vector) flux on the Neumann boundary.

The tensor (3) can be diagonalized by the transition to total electron energy $\varepsilon = u - e\varphi(\boldsymbol{r})$ as the independent variable ($\varphi(\boldsymbol{r})$ is the electrostatic potential, $\boldsymbol{E} = -\nabla\varphi$). The diagonalization results in the diffusion tensor of the form

$$D_\varepsilon = D_r \begin{pmatrix} 1 & 0 & 0 & 0 \\ 0 & 1 & 0 & 0 \\ 0 & 0 & 1 & 0 \\ 0 & 0 & 0 & 0 \end{pmatrix} \qquad (5)$$

The total energy approach has been previously used for solving many kinetic problems in gas discharge physics. [20,21] Using total energy is convenient for analyzing the mechanisms of EEDF formation and obtaining approximate analytical solutions because the cross derivatives in the FP equation (1) vanish, and the kinetic equation has the simplest form. However, the region of the phase space in which this equation has to be solved has a curvilinear boundary $\varepsilon = -e\varphi(\boldsymbol{r})$. This presents difficulties for numerical calculations, primarily since the shape of the boundary is determined by an unknown electric potential $\varphi(\boldsymbol{r})$.[22]

In the present study, we introduce the concept of surface diffusion for solving the FP equation (1) in phase space using kinetic energy as an independent variable. Anisotropic diffusion appears in various scientific fields, including magnetized plasmas, flows in porous media, image processing, etc. Surface diffusion is a particular case of strongly anisotropic diffusion where the diffusion coefficient normal to the surface is zero.[23] The computational challenges of solving the full-tensor diffusion problems and an example of a surface diffusion problem are described in Appendix.

To illustrate the surface diffusion concept for the FP equation (1), consider the problem of electron kinetics in a spatially uniform electric field $E_0$. We neglect the energy loss in elastic collisions ($\Gamma_u = 0$) and take into account the excitation of only one lowest excited state of atoms with energy $\varepsilon_1$. The corresponding collision operator is described by

$$C_0 = -\nu^*(u)f_0(u) + \frac{\sqrt{u+\varepsilon_1}}{\sqrt{u}}\nu^*(u+\varepsilon_1)f_0(u+\varepsilon_1) \qquad (6)$$

where $\nu^*(u)$ is the corresponding collision frequency (which vanishes at $u < \varepsilon_1$). The first term in (6) describes a sink of high-energy electrons as a result of their energy loss. And the second term describes the appearance of low-energy electrons after collisions. The diffusion tensor has the form:

$$D = D_r \begin{pmatrix} 1 & E_0 \\ E_0 & E_0^2 \end{pmatrix} \qquad (7)$$

Figure 1 shows a steady-state solution of the FP equation (1) with the boundary condition at $x=0$ corresponding to electron injection with a Gaussian distribution from the cathode. The electron mean free path $\lambda = v/\nu$ is assumed to be small compared to the gap length $L$. The injected electrons diffuse over surfaces of constant total energy $\varepsilon = u - eE_0 x$, until they experience discrete kinetic energy losses (vertical jumps) due to inelastic collisions with atoms. As a result, a periodic structure is formed in the phase space $(x,u)$ with a spatial period $\lambda_\varepsilon = \varepsilon_1/(eE_0)$. The electron energy relaxation length in the inelastic energy range $\lambda^*$ is assumed to be small compared to $\lambda_\varepsilon$, so that the low-energy electrons appear at energies much lower than $\varepsilon_1$. These requirements correspond to the conditions of the famous Franck-Hertz experiments.[24]

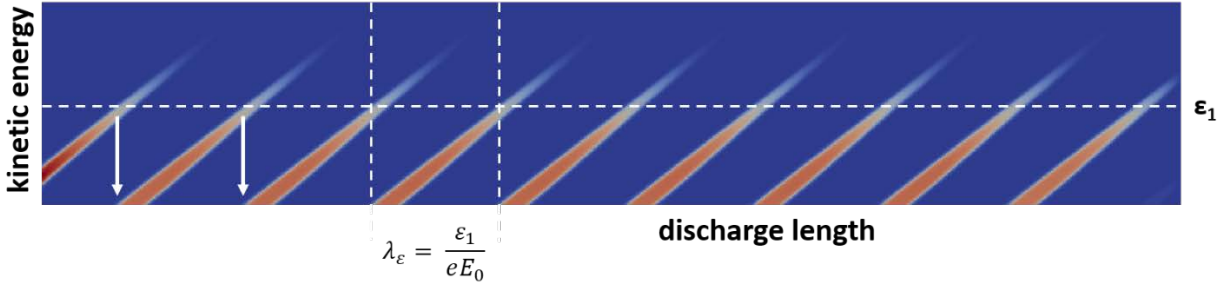

*Figure 1: A periodic structure in phase space formed due to surface diffusion and energy jumps of electrons in a spatially uniform electric field.*

In the one-level model, which takes into account only the lowest state of atoms, the electrons remember their energy distribution at the cathode, and the EEDF is perfectly periodic in space (Figure 1). Excitations of several atomic levels with different energy thresholds result in a spatial relaxation process associated with damping striations. In this process, the electrons forget their energy distribution at the cathode.[25] The ionization of atoms by electron impact and the generation of secondary electrons further decrease the spatial relaxation length.[10] The electron energy loss in elastic collisions with atoms provides yet another relaxation mechanism, which must be considered when this loss contributes to the overall electron energy balance in the plasma of PC.

## Self-Consistent Hybrid Discharge Model

We have developed full-tensor diffusion solvers using COMSOL[26] and Basilisk C.[27] Our COMSOL solver follows the works [28,29]. The Basilisk solver extends the existing open-source diffusion solver available from the default installation by adding off-diagonal tensor components.

In our current hybrid discharge model, the electrons are described by the kinetic equation in a 2d $(x, u)$ phase space:

$$\frac{\partial f_0}{\partial t} - \nabla \cdot (\mathbf{D}_2 \nabla f_0) - \frac{1}{\sqrt{u}} \frac{\partial}{\partial u}\left(\sqrt{u} \Gamma_u\right) = C_0 \qquad (8)$$

where the diffusion tensor is given by

$$\mathbf{D}_2 = D_r \begin{pmatrix} 1 & E(x,t) \\ E(x,t) & E(x,t)^2 \end{pmatrix} \qquad (9)$$

In addition to incorporating the excitation of the lowest excited level of atoms (6), we include the direct ionization by electron impact in the form:[8]

$$C_{ion} = -v^{ion}(u)f_0(u) + 4\frac{\sqrt{2u+\varepsilon_{ion}}}{\sqrt{u}} v^{ion}(2u + \varepsilon_{ion})f_0(2u + \varepsilon_{ion}) \quad (10)$$

where $v^{ion}(u)$ is the ionization frequency, and $\varepsilon_{ion}$ is the ionization threshold. This ionization model assumes that the kinetic energy is evenly distributed between the primary and secondary electrons after an ionization event. The electron loss to the wall of the cylindrical tube of radius $R$ is included in the form:[5,8]

$$C_{wall} = -\frac{1}{3}\left(\frac{v}{R}\right) f_0(u)\Theta(u - \Phi_w) \quad (11)$$

where $\Phi_w(x)$ is the wall potential with respect to plasma, and $\Theta(x)$ is the step function.

Ions are described using a drift-diffusion model:

$$\frac{\partial n_i}{\partial t} + \frac{\partial}{\partial x}\left(\mu_i n_i E(x) - D_i \frac{\partial n_i}{\partial x}\right) = I - \frac{n_i}{\tau} \quad (12)$$

where $\mu_i$ and $D_i$ are the ion mobility and diffusion coefficients, and $I$ is the ionization rate by electron impact. The term $\frac{n_i}{\tau}$ describes the radial ambipolar loss, where $\tau = (R/2.4)^2/D_a$ is the effective time of the radial ambipolar diffusion to the wall, $D_a$ is the ambipolar diffusion coefficient. The electric field is calculated from Poisson's equation, assuming a grounded anode and a voltage $U$ applied to the cathode.

The equations (8-12) must be supplemented by the boundary conditions to specify the discharge properties. Most experimental studies of DC discharges were obtained for cold cathodes with secondary electron emission by ion impact. However, hot cathodes and a plasma rf cathode [30] have also been used, and plasma stratification processes appeared not very sensitive to the cathode processes. To simplify simulations of the cathode region, we use the hot cathode model in the present paper. Electrons were injected from the cathode with a Maxwellian distribution and a prescribed electron flux, which was assumed to be lower than the electron flux in the PC to provide some electron multiplication in the cathode region.

The energy range $u_{max}$ for the electron kinetic equation is selected about 2-3 $\varepsilon_1$, and the boundary condition at $u_{max}$ is specified as $f_0(u = u_{max}) = 0$.

The boundary condition at $u = 0$ corresponds to the absence of electron flux from the boundary,

$$\boldsymbol{D_2}\nabla f_0 \cdot \hat{n} = 0 \quad (13)$$

where $\hat{n}$ denotes the unit normal. Equation (13) is a more specific version of Eq. (4).

By imposing that $f_0$ be continuously differentiable and bounded, we obtain from (13):

$$\frac{\partial f_0}{\partial x} - E \frac{\partial f_0}{\partial u} \to 0. \tag{14}$$

as $u \to 0$, assuming that $\sqrt{u} C_0 \to 0$.

The boundary condition at the cathode $(x = 0)$ corresponds to electron emission with a Maxwellian distribution $g(u)$ and a specified current density $\Gamma_e$:[28]

$$-\sqrt{u} D_r \left( \frac{\partial f_0}{\partial x} - E \frac{\partial f_0}{\partial u} \right) = \frac{u}{2m} f_0 - \Gamma_e \frac{m}{4\pi} g(u) \tag{15}$$

The boundary condition at the anode $(x = L)$ is:

$$-\sqrt{u} D_r \left( \frac{\partial f_0}{\partial x} - E \frac{\partial f_0}{\partial u} \right) = \frac{u}{2m} f_0 \tag{16}$$

The boundary conditions for the ions are:[28]

$$\pm \left( \mu_i n_i E - D_i \frac{\partial n_i}{\partial x} \right) = \frac{n_i v_i^{th}}{4} \pm \Theta(\pm E) E \mu_i n_i \tag{17}$$

where $v^{th}$ is the ion thermal speed, and different signs correspond to the cathode (-) and anode (+).

The initial EEDF is Maxwellian, and the ion density is equal to the electron density.

The Basilisk framework uses a finite volume discretization scheme. It has explicit and implicit capabilities for tensor diffusion simulations, with limiters available in the explicit version. Our studies found that the superbee limiter helps eliminate the unphysical negative values in the distribution function without artificially adding diffusion. The implicit version uses a multigrid relaxation (iterative) method at each time step. The drift-diffusion equations for the ion transport and Poisson equation are solved implicitly using a tridiagonal (Thomas) algorithm. The 1d1u kinetic equation for electrons and 1d equations for ions and the electric field are solved sequentially to obtain a self-consistent solution for the discharge dynamics. In the explicit solver, the fast electron kinetics limits the time step.

The COMSOL solver is based on a finite element framework. It also has explicit and implicit capabilities, although only the implicit (BDF) time-stepping method was employed in the present studies. A direct (MUMPS) solver is employed at each time step for all quantities. The negative values of $f_0$ and $n_i$ are avoided by utilizing the logarithm formulation of the ion drift-diffusion and electron kinetic equations. The coupled equations of different dimensionality are solved simultaneously.

## Results of Simulations

We have performed self-consistent 2-level (excitation and ionization) simulations of an Argon plasma in a discharge tube of radius 1 and 0.5 cm, gas pressure 0.4 Torr, for different lengths from 5 cm to 40 cm. Two models of Argon plasma have been used: a 2-level model (excitation of the first excited state with the threshold $\varepsilon_1 = 11.55\ eV$ and direct ionization with the threshold $\varepsilon_{ion} = 15.8\ eV$) and the full chemistry model, which included stepwise and Penning ionization. In the latter case, we considered diffusion of metastable atoms with their deactivation at the discharge tube walls and electrodes.

The current in the PC was calculated assuming a Bessel radial distribution of plasma density and was controlled by varying the electron flux injected from the cathode. The potential drop over striation length $\Delta\varphi_\Lambda$ and the average electric field $\langle E \rangle$ can be calculated as:

$$\Delta\varphi_\Lambda = \int_0^\Lambda dx E(x,t) = v_s \int_0^T dt E(x,t) = \langle E \rangle \Lambda \tag{18}$$

Here, $v_s$ denotes the speed of the ionization wave, $T = 1/v_s$ is its period, and $v_s$ is the frequency. The average electron density $\langle n_e \rangle$ in the PC was calculated the same way as $\langle E \rangle$, by monitoring the time variation of $n_e(t)$ at a fixed point.

Table 1 illustrates the selected cases discussed below. Cases 1 and 2 are for $p = 1$ Torr, while the rest are for $p = 0.4$ Torr. Also, cases 9 and 10 are for $R = 0.5$ cm, all others are for $R = 1$ cm.

Table 1. Details of selected simulation cases

| # | Gap (cm) | Voltage (V) | Emiss. Current (µA) | PC Current (mA) | Chem. model | $\langle n_e \rangle$ ($10^8$ cm$^{-3}$) | $\langle E \rangle$ (V/cm) | $\Delta\varphi_\Lambda$ (V) | $\Lambda$ (cm) | Freq. $v_s$ (kHz) |
|---|---|---|---|---|---|---|---|---|---|---|
| 1 | 5 | 100 | 5e-5 | 6.1e-8 | 2-lev. | 2.3x10$^{-7}$ | 20 | 11 | 0.57 | 0 |
| 2 | 5 | 100 | 5e-2 | 4.8e-5 | 2-lev | 2.0x10$^{-4}$ | 20 | 8.4 | 0.42 | 0 |
| 3 | 10 | 130 | 5 | 0.016 | 2-lev | 0.078 | 6.1 | 12 | 1.9 | 0 |
| 4 | 10 | 130 | 500 | 0.98 | 2-lev | 5.4 | 5.2 | 12 | 2.4 | 37 |
| 5 | 20 | 160 | 50 | 1.3 | full | 35 | 1.7 | 8.0 | 4.7 | 0.33 |
| 6 | 20 | 160 | 500 | 0.99 | 2-lev | 6.9 | 4.0 | 15 | 3.8 | 35 |
| 7 | 20 | 200 | 500 | 0.56 | full | 14 | 1.8 | 10 | 5.7 | 0.29 |
| 8 | 40 | 200 | 35 | 0.26 | full | 4.5 | 1.9 | 10 | 5.0 | 0.23 |
| 9 | 10 | 130 | 125 | 0.29 | 2-lev | 4.9 | 6.5 | 17 | 2.6 | 81 |
| 10 | 20 | 160 | 125 | 2.5 | full | 80 | 3.1 | 6.8 | 2.2 | 0.8 |

## Standing striations

Standing striations were often observed in experiments for Townsend discharges in noble gases in a specific range of gas pressures. We obtained them in our simulations at low discharge currents. Figure 2 illustrates an example of Townsend discharge for the 5 cm gap and 100 V voltage between electrodes, and discharge current 50 pA. The electric field between the electrodes is almost uniform and has a value of 20 V/cm.

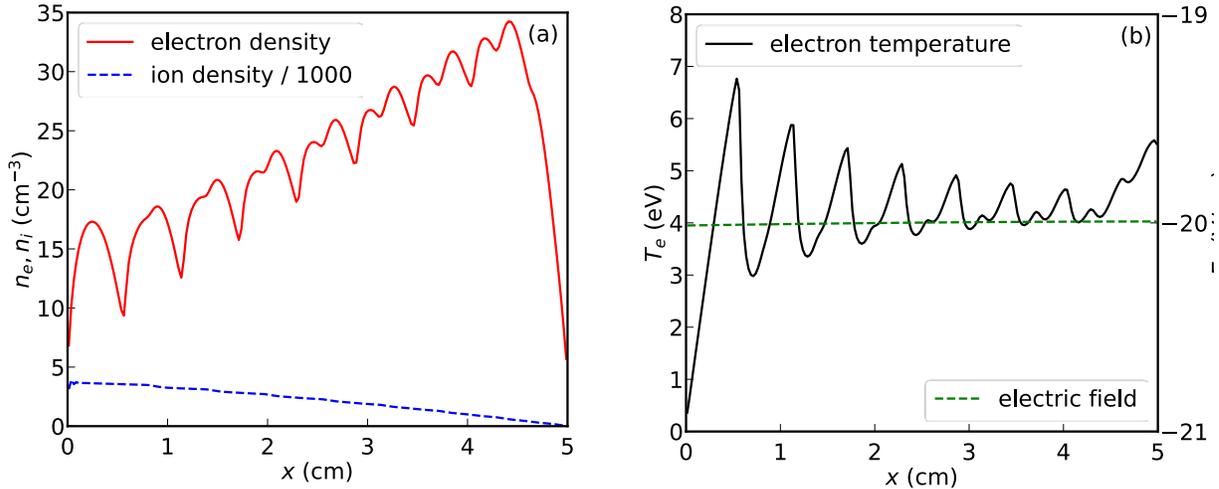

*Figure 2: Electron and ion density (a) and electron temperature and electric field (b) in Townsend discharge at current of 50 pA (case 1 in Table 1).*

Figure 2 shows spatial distributions of the electron and ion densities and electron temperature obtained in a steady state. The electron density increases towards the anode due to electron multiplication and has double-peaked structures due to the 2-level model.

The discharge transits into a subnormal regime with increasing current, and standing striations become heavily damped as shown in Figure 3.

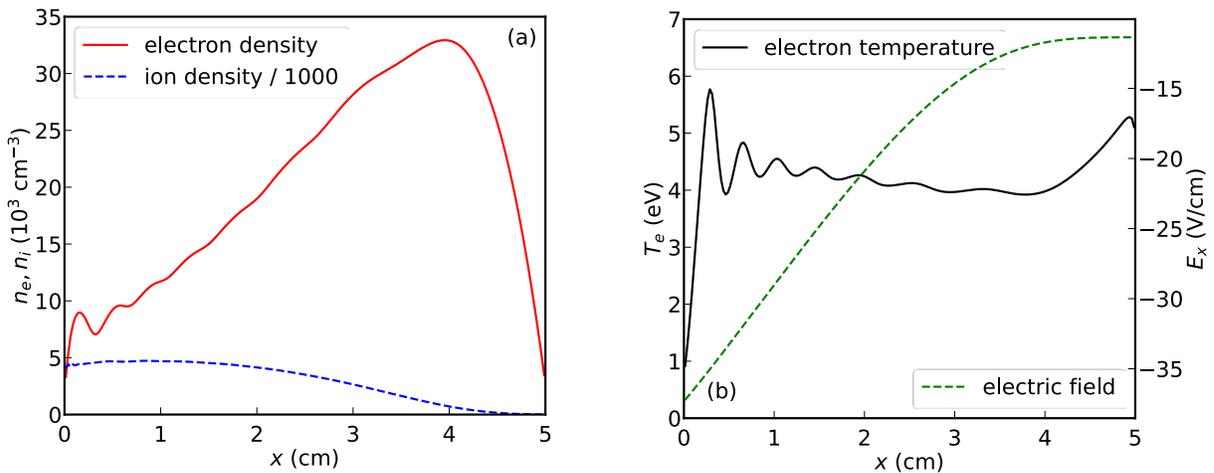

*Figure 3: Electron and ion density (a) and electron temperature and electric field (b) in a subnormal discharge at current 50 nA (case 2 in Table 1).*

Standing striations have also been observed in our simulation at higher currents for the Debye length in the range $R \leq r_D < L$. Figure 4 shows an example of simulations for the gap length of 10 cm, $U = 130$ V, and current 5 µA. The spatial distributions of electron and ion densities (a) demonstrate notable deviations from quasi-neutrality in the cathode sheath and the PC with weakly damped standing striations. The electric field (b) is highly non-uniform but only slightly perturbed by the striations in the positive column. The electric field does not reverse sign even in the cathode region. The electron temperature is strongly modulated in the positive column and increases in both the cathode sheath and near-anode plasma, as expected in DC discharges under these conditions.

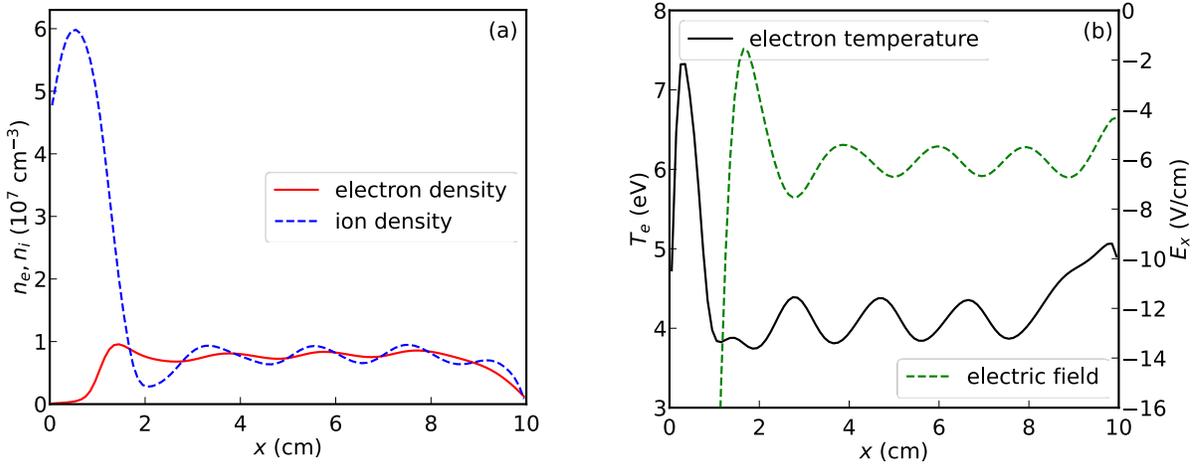

*Figure 4: Glow discharge with standing striations in the positive column at current 5 µA (case 3 in Table 1).*

## Moving striations

Moving striation appeared in our simulations at currents exceeding a critical value. The critical currents are about 10 µA and depend on the chemistry model used. The moving striations first appear at the anode and travel towards the cathode with speeds ranging from 10 to 1300 m/s depending on discharge conditions and the chemistry model. The speed of ion-guided waves exceeds the speed of metastable-guided waves by two orders of magnitude.

Figure 5 shows two-dimensional *(x,t)* contours of electron density in moving striations for the ion-guided waves (2-level model) and metastable-guided waves (full-chemistry model) for the 10 cm and 20 cm gaps, respectively. The 2-level model reaches a dynamic steady-state much faster (~0.1 ms) than the full-chemistry model (~3 ms). Interestingly, the full chemistry model shows a transient phase (< 2 ms) which behaves similarly to the 2-level model (not resolved in the Figure). The direct ionization processes initially control the system. At longer times, with the growth of metastable population, stepwise ionization becomes dominant.

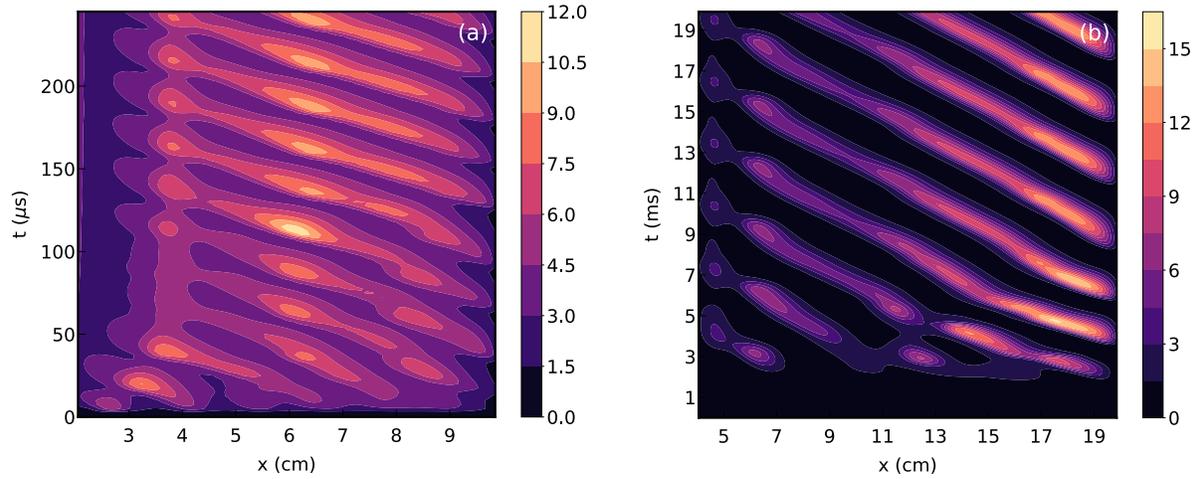

*Figure 5: Contours of constant electron density in space and time for the 2-level model in units of $10^8$ cm$^{-3}$ (a) and full chemistry model in units of $10^9$ cm$^{-3}$ (b).*

Figure 6 shows the instantaneous axial structure of the discharge in the presence of moving striations for a 20 cm gap using the 2-level model. The cathode sheath and the negative glow with a prominent plasma density peak are visible. The stratified plasma in the positive column is quasineutral, and the striations are highly nonlinear. A movie illustrating the formation of this structure and the dynamics of plasma stratification, including the initial appearance of moving striations near the anode and their propagation towards the cathode, is available on the journal website.

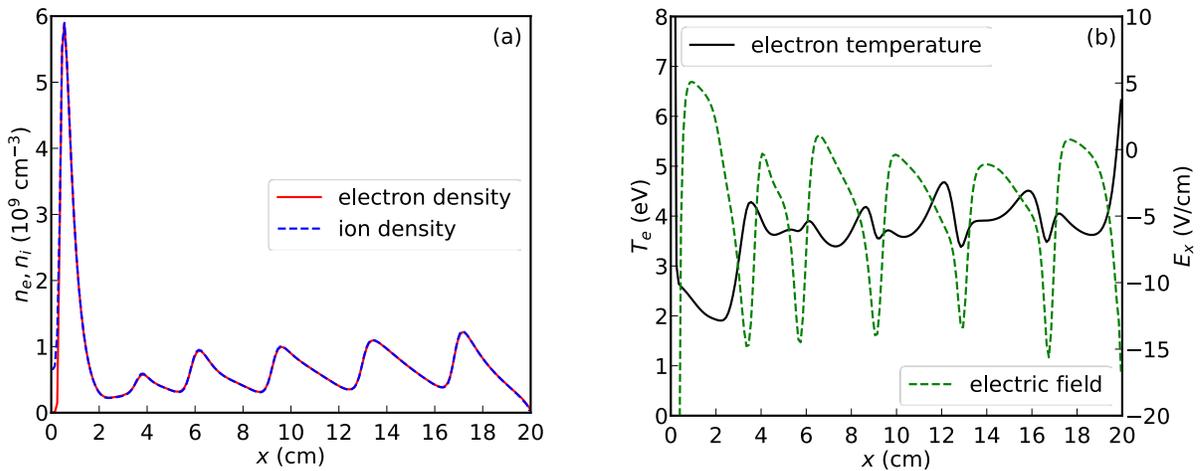

*Figure 6: Instantaneous spatial distributions of charged particle densities (a), electric field, and electron temperature (b) for current 0.5 mA and voltage 160 V (case 6 in Table 1).*

Figure 7 shows the temporal variation of plasma density and electric field in the positive column for the 2-level model. It is seen that moving striations appear at the time about 10 µs, which is of the order of the radial ambipolar diffusion time at these conditions. The electric field does not change its sign (in field reversals) in the positive column but changes the sign at the anode once during the striation period (see below).

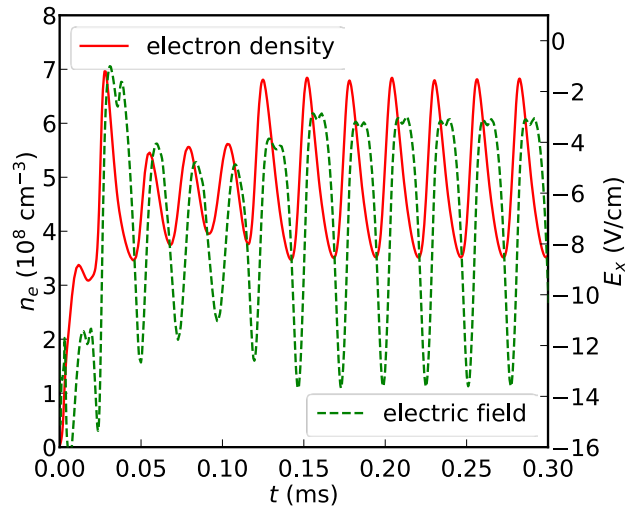

*Figure 7: Time dependence of plasma density and electric field in the positive column for 500 µA, 2-level model (case 4 in Table 1).*

Figure 8 shows results for a similar full-chemistry case. Now, the electric field changes its sign in the positive column (PC). The electric field reversals are also observed at the anode, as in the 2-level model. The plasma stratification with the full chemistry model occurs in two stages. First, at times up to about 0.4 ms (when the 2-level case has already reached a steady-state), the full-chemistry case evolves irregularly and the electron density gradually increases by an order of magnitude (not resolved in Figure 8). After about 1 ms, the ionization wave settles into a much lower frequency and velocity.

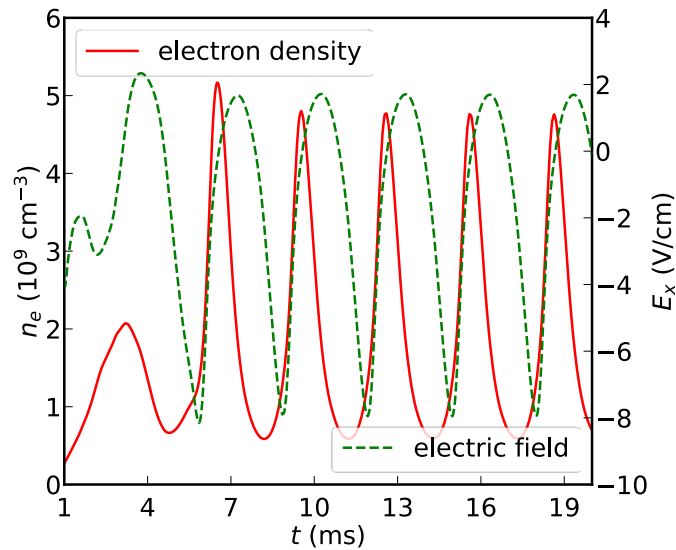

*Figure 8: Time dependence of plasma density and electric field in the positive column for 20 cm, 160 V, 50 µA, full-chemistry model (case 5 in Table 1).*

In the full chemistry model, initial discharge development is controlled by direct ionization, and the striations are controlled by ion drift. After about 0.1 ms, when the metastable atoms accumulate to high levels, stepwise and Penning ionization become dominant. The striation frequency drops by two orders of magnitude as the striation motion begins to be controlled by slow metastable diffusion. Such transition is accompanied by a factor of two drop of the (time-averaged) electric field in the PC. In the presence of substantial stepwise and Penning ionization, lower electric fields are required to sustain the plasma controlled by radial ambipolar losses.

The electric field reversals at the anode obtained in our simulation with both 2-level and full-chemistry models (shown in Figure 9) indicate oscillations between the ion and electron anode sheaths previously observed in experiments.[31] It appears that the anode region controls the type of moving striations and their amplitude for the considered discharge conditions.

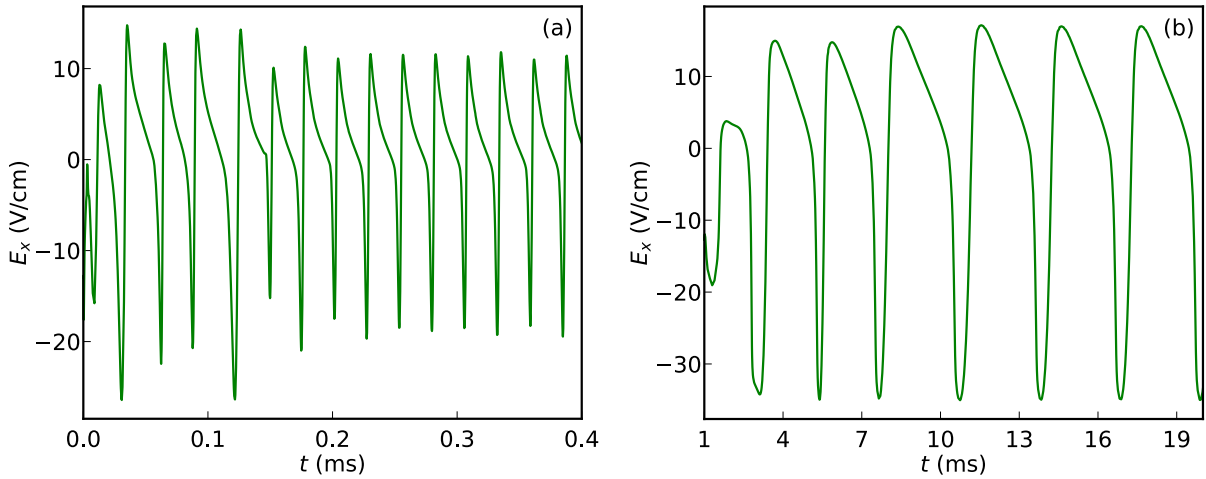

*Figure 9: Time dependence of the electric field at the anode for 2-level model (a) and full chemistry model (b) (cases 4 and 7 in Table 1).*

Figure 10 shows instantaneous distributions of plasma parameters over striation length for the 2-level model. The potential drop over striation length is about 16 eV, which indicates the *s*-waves. The electric field does not change its sign. The reason for the ionization wave propagation towards the cathode is seen in Figure 10. The ionization rate is shifted towards the cathode with respect to plasma density, which forces the ionization wave to propagate towards the cathode.

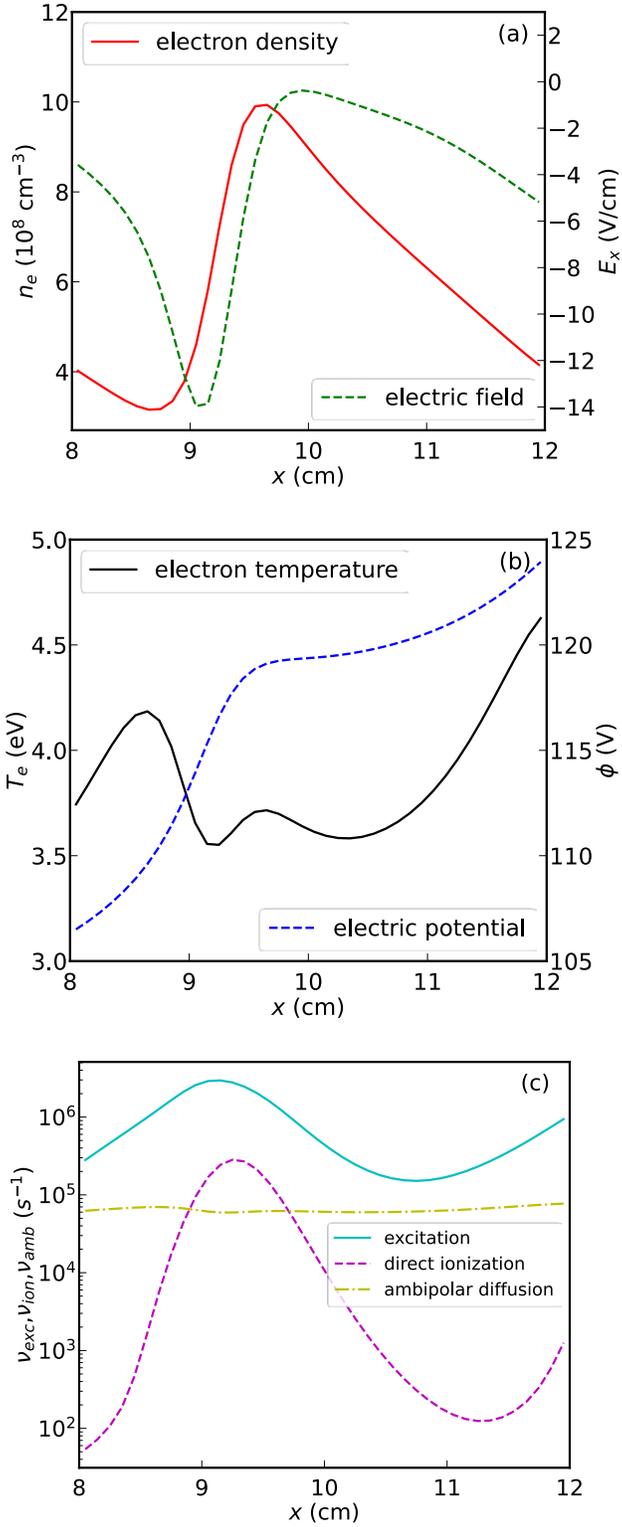

*Figure 10: Spatial distribution of electron density and electric field (a), electron temperature and electric potential (b), and ionization and radial ambipolar loss frequencies (c) over striation length for current 0.5 mA, in the 2-level model (t = 0.25 ms, case 6 in Table 1).*

Figure 11 shows the spatial distribution of the plasma parameters over striation length that could be compared with the 2-level results shown in Figure 10. The first key difference is that the electric field changes sign forming electrostatic traps for the slowest electrons. The second key difference is that the potential drop over striation length is about 8 eV, which indicates that these striations are of the *p*-type. Third, the main ionization channel appears to be Penning ionization, followed by stepwise ionization. As a result, according to the full-chemistry model, the average electric field in the PC is two times lower compared to the 2-level model, which includes only direct ionization (see Table 1).

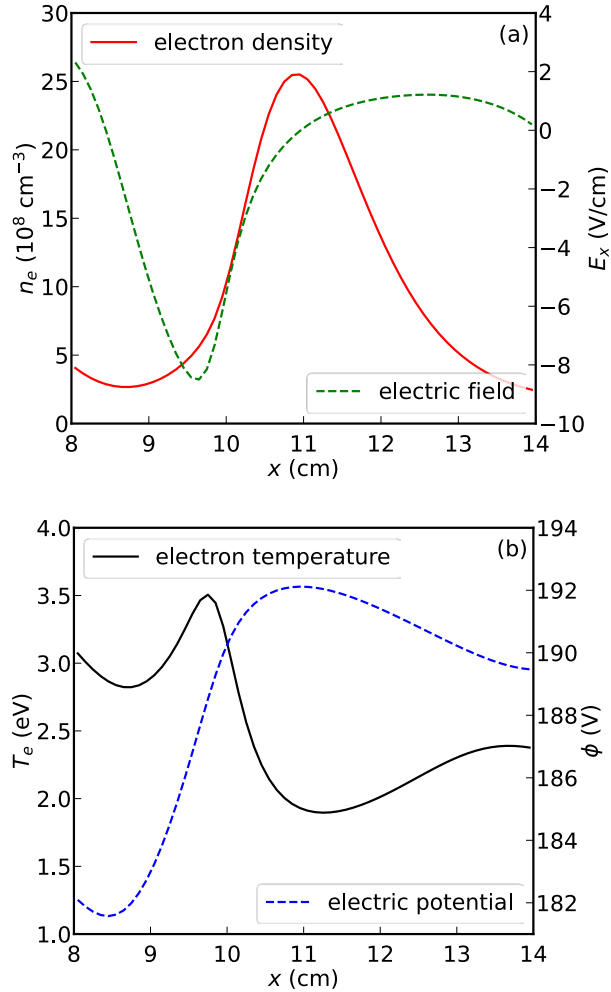

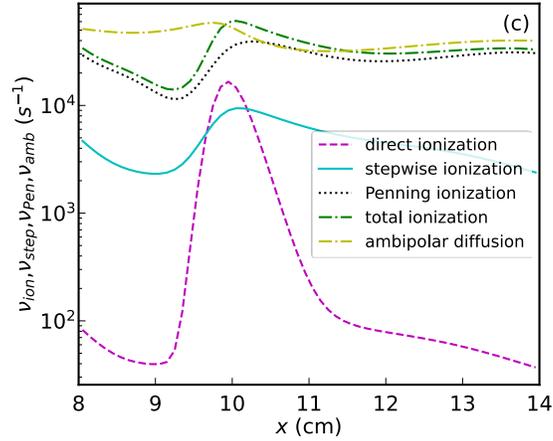

*Figure 11: Spatial distribution of electron density and electric field (a), electron temperature and electric potential (b), and ionization and radial ambipolar loss frequencies (c) over striation length for current 0.5 mA for the full chemistry model (t = 20 ms, case 7 in Table 1).*

Figure 12 shows the calculated contours of the EEDFs for *s* and *p* striations. These contours are substantially different from those shown in Figure 1 for a spatially uniform electric field. The structure of EEDFs in strongly modulated electric fields indicates the presence of two regions over striation length. Local electron heating (diffusion in energy) occurs in the areas of the strong electric field where the terms proportional to *E* in the kinetic equation (1) dominate over spatial diffusion. Outside of the heating region, the spatial diffusion of electrons dominates over diffusion in energy. Overall, the spatial and energy diffusion balance each other over striation length to ensure the potential drop over striation length satisfies Novak's law. This behavior is consistent with the qualitative model of nonlinear kinetic striations proposed by Tsendin.[32]

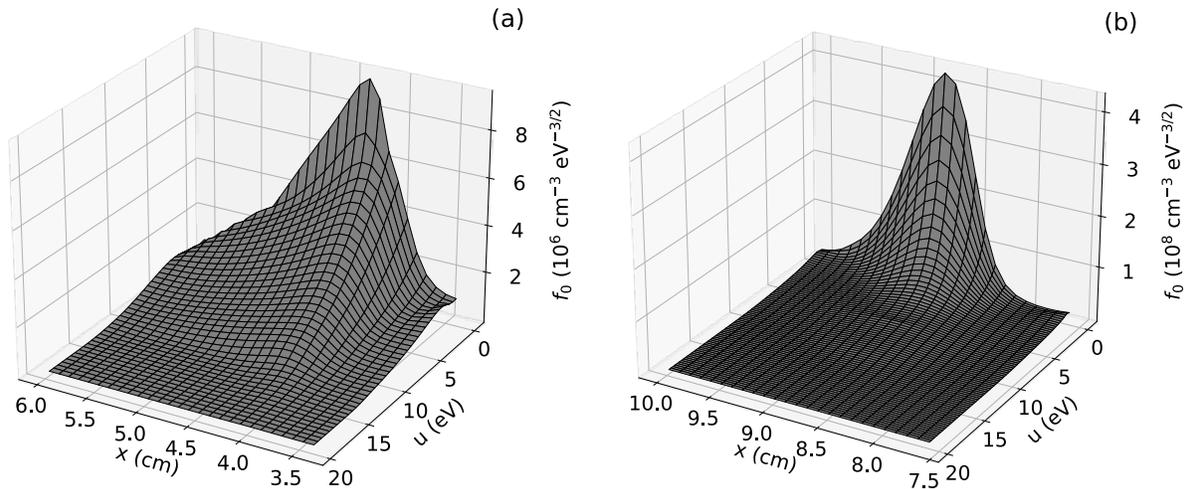

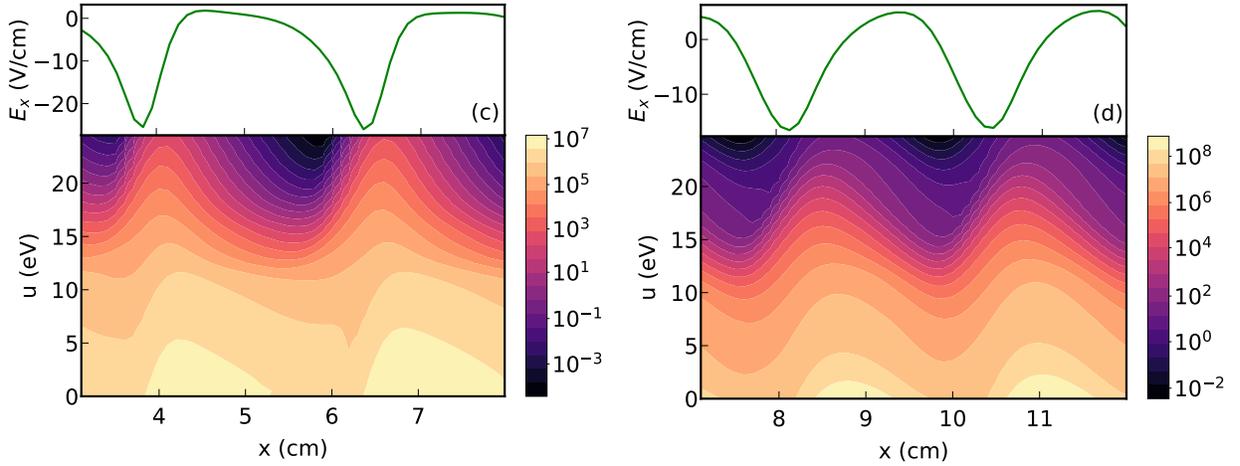

*Figure 12: Instantaneous EEDFs for s-striations (a,c) (case 9 from Table 1) and p striations (b,d) (case 10 from Table 1) in linear and log scales. Inserted are the electric field profiles.*

The results shown in Figure 12 indicate substantial differences with previous calculations [35] for *s* and *p*-striations in *prescribed* weakly-modulated electric fields. These differences are expected because our *s* and *p* striation*s* in *short discharges* are highly nonlinear. Furthermore, the previous calculations, [35] which were performed for *long gaps* to study the electron bunching effect due to energy loss in elastic collisions, neglected contribution of ionization processes in EEDF relaxation. The electron energy loss in elastic collisions was negligible for short discharges under our conditions. We can conclude that for short discharges at low gas pressures, ionization processes in the cathode region and near-anode phenomena must be responsible for selecting the wave type observed in experiments.

## Discussion

We have introduced the concept of surface diffusion in phase space *(r,u)* to solve the Fokker-Planck (FP) kinetic equation for electrons. According to this concept, electrons diffuse along surfaces of constant total energy $\varepsilon = u - e\varphi(r)$ and experience energy jumps in inelastic collisions with atoms. The surface diffusion concept appears more convenient for the numerical solution of the FP kinetic equation than the *total energy* approach previously used for the theoretical analysis of electron kinetics. Although the total energy approach simplifies the kinetic equation by diagonalizing the tensor diffusion coefficient, the region of the phase space in which the equation has to be solved is bounded by a surface $\varepsilon = -e\varphi(r)$, which is determined by an unknown electric potential $\varphi(r)$. The potential has to be found self-consistently, which makes numerical calculations more complicated. [22] Using kinetic energy as an independent variable simplifies the numerical solution of the FP kinetic equation for electrons in collisional plasmas.

We have obtained numerical solutions of the 1d1u FP kinetic equation for electrons by the finite volume and finite element methods. The kinetic solver for electrons was coupled to ion transport and Poisson solver for the electric field. Using this method, we studied the dynamics of stratification in Townsend and glow discharges in Argon gas. Standing striations appeared in Townsend discharges at low discharge currents and in the cathode region of glow discharges at larger currents. Moving striations appeared at discharge currents exceeding a critical value. We believe that our work helped identify the discharge currents that correspond to the left boundary of the striation appearance with respect to *i/R* in the experimental (*i/R, pR*) maps.

We have studied the dynamics of discharge development using the -level and full-chemistry models. Simulations showed that moving striations originated in the anode region and propagated towards the cathode in both cases. Two types of moving striations have been identified resembling the *s* and *p* striations previously observed in the experiments. It appears that the anode region plays a vital role in controlling the type of moving striations and their amplitude in the positive column plasma.

The question about standing and moving striations in DC discharges has remained one of the key questions in this field.[20] Both types of striations have been observed experimentally in DC discharges of noble gases.[33] Standing striations often appeared near perturbations of plasma density in the positive column. We obtained both standing and moving striations with our code and shed new light on the nature of plasma stratification in noble gases. In our 2-level model (excitation and ionization), standing striations were formed in Townsend, subnormal, and glow discharges. Standing striations in glow discharges appeared between the negative glow (NG) and PC.

The possibility of the complicated structure of the Faraday dark space was previously discussed in Ref. [34]. This region between the NG and PC deserves further experimental and theoretical studies. In the cathode region, spatial relaxation and electron multiplication processes make electrons forget their history. As a result, the plasma of PC becomes autonomous and not sensitive to the cathode emission processes. Standing striations in the region between the NG and PC contain valuable information about nonlocal electron kinetics, ion transport, and chemical processes. While the electron energy loss in elastic collisions with atoms (the $\Gamma_u$ term in Eq. (1)) was included in our simulations, its effect on the EEDF formation was negligible compared to relaxation processes due to electron multiplication in the cathode region. When the elastic collisions determine the electron energy balance in the PC, they must be considered in the transition region between the NG and PC.

In the case of moving striations, which appear near the anode and propagate towards the cathode, electric field perturbations support density perturbations to cause non-damped waves. The spatial shift of the ionization rate towards the cathode with respect to maximal plasma density makes the ionization wave move towards the cathode. Our simulations produced moving striations resembling the *s* and *p* striations in short discharges and clarified the role of electrodes in plasma stratification. We have observed that our simulation results are not sensitive to the electron loss to the wall described by Eq. (11). In fact, completely neglecting the electron loss term resulted in minor changes of simulation results under our discharge conditions, in agreement with conclusions drawn in Ref. [8]. This indicates an important difference between plasma stratification in short and long discharges, which can be obtained by imposing periodic boundary conditions to neglect effects of near-electrode regions. In the latter case, the ionization and loss of electrons must be exactly balanced over the striation length. Further studies of electrode effect on plasma stratification and comparison of calculated EEDFs with experimental observations using Langmuir probes [30] are planned in future work.

Future studies of *s*, *p*, and *r* striations in noble gases under specific discharge conditions are planned. These striations have been associated with electron bunching and nonlinear resonances.[35] Our results indicate that moving striations of the *s* and *p*-type are generated near the anode and become highly nonlinear in short discharges. The electron bunching effects due to energy loss in elastic collisions are not crucial for the short discharges under studied conditions, as they occur over much longer distances. The nonlinear effects associated with self-supported perturbations of plasma density, ionization rate, and the electric field appear to be the primary reason for selecting a particular type of kinetic striations in short tubes. Detailed kinetic studies of the anode region effects on plasma stratification in noble gases could be a subject of future work.

We have used a discharge model with hot (thermo-emitting) cathodes for self-consistent simulations of DC glow discharges. It was previously shown in simulations [29] that the basic structure of the cathode region in DC discharges with cold and hot cathodes is similar as long as the electron multiplication factor in the NG exceed a factor about two. The hot cathode model results in a lower potential drop in the cathode sheath, simplifying the discharge simulations reported in the present paper. The cathode region was shown to play an important role in the dynamics of plasma stratification in short discharges under studied conditions. Clearly, the structure of the cathode region and its effects on the striations in PC depend on discharge conditions and deserve further studies.

Both COMSOL and Basilisk codes appear suitable for hybrid plasma simulations. These codes have different discretization schemes and enable both implicit and explicit FP solvers for electrons. In our Basilisk simulations, we have attempted to artificially increase the ion mobility (100X) to speed up convergence to steady-state and plan to explore this avenue further. The implicit version of the FP solver produced a 187X speed up but yielded artificially damped striations. The explicit code with a superbee flux limiter gave qualitatively like COMSOL results but at a high computational cost. Our studies highlight the need for an implicit kinetic FP solver capable of adequately handling the negative EEDF values. Negative (unphysical) distribution function values occur in this setting due to the presence of non-zero cross diffusion terms (off diagonal elements in the diffusion tensor). At the same time, the high memory demands (~10GB per run) of the COMSOL implicit coupled solver compared to lower memory demands (~1MB) of the Basilisk solver makes Basilisk more attractive for future multi-dimensional hybrid plasma simulations.

## Conclusions

We developed a self-consistent hybrid model of ionization waves (striations) in low-current DC discharges in noble gases using a Fokker-Planck kinetic solver for electrons, a drift-diffusion model for ions, and a Poisson's solver for the electric field. The FP kinetic equation for the EEDF has the form of a full-tensor anisotropic diffusion in phase space. We have introduced the concept of *electron diffusion over a surface of total energy in phase space* to interpret and validate the tensor diffusion solver. The first self-consistent simulation of the "kinetic" striations using the nonlocal Fokker-Planck kinetic equation-based approach was described.

We have studied the formation of standing and moving striations in low-current DC discharges using the 2-level (excitation-ionization) model and a "full chemistry" model, which includes stepwise and Penning ionization. Standing striations appeared in Townsend and subnormal discharges. They also appeared in the transition region between the negative glow and positive column in glow discharges. We have obtained two types of moving striations for the 2-level and full chemistry models, which resemble the *s* and *p* striations previously observed in the experiments. The electric field reversals at the anode have been observed in our simulation. It appears that processes in the anode region control the type of moving striations and their amplitude. A detailed comparison of model predictions with available experimental observations is planned in future work.

## Appendix: Diffusion over a ring

Considerable literature has been devoted to solving anisotropic diffusion problems.[36] Diffusion over a surface is a particular case of strongly anisotropic diffusion with a zero-diffusion coefficient in the direction normal to the surface. A sample of surface diffusion is the diffusion over a ring, which is described in a 2D Cartesian coordinate system by the diffusion tensor:[37],[38]

$$D = \frac{\kappa}{x^2+y^2}\begin{pmatrix} -y^2 & xy \\ xy & -x^2 \end{pmatrix} \quad (19)$$

In the cylindrical coordinate system, $x = r\cos(\theta)$ and $y = r\sin(\theta)$, the diffusion tensor becomes diagonal with only one non-zero component. The diffusion occurs only over angle $\theta$ and is described by the Laplace-Beltrami operator:[38]

$$\frac{\partial f}{\partial t} - \frac{\kappa}{r^2}\frac{\partial^2 f}{\partial \theta^2} = 0 \quad (20)$$

Thus, the off-axis components of the diffusion tensor (18) ensure the absence of radial diffusion. The solution of (19) must be periodic with respect to $\theta$ with a period $2\pi$.

The numerical solution of anisotropic diffusion problems could yield anisotropy-dependent convergence, introduce large perpendicular errors, and a loss of positivity near high gradients. The ring diffusion is an excellent test to check the properties of the anisotropic diffusion scheme because field lines make all possible angles with respect to the Cartesian grid.

Consider an initial condition for the ring diffusion problem in the form of a Gaussian peak given by

$$f(t=0) = \frac{1}{\delta\sqrt{\pi}} \exp\left(-\frac{(x-x_0)^2+y^2}{\delta^2}\right). \quad (21)$$

The Gaussian peak corresponds to a delta function for $\delta \to 0$. The solution must be symmetric, which implies the Neumann boundary condition for Eq. (19) at $\theta = \pi$.

The ring diffusion problem described by Eq. (19) has a known analytical solution:[23]

$$f(t,\theta) = \frac{a_0}{2} + \sum_{n>1} a_n \exp\left(-\frac{n^2 \kappa t}{r^2}\right) \cos(n\theta) \tag{22}$$

where

$$a_n = \frac{1}{\pi}\int_{-\pi}^{\pi} f(\theta) \cos(n\theta)\, d\theta = \frac{2}{\pi}\int_{0}^{\pi} f(\theta) \cos(n\theta)\, d\theta \tag{23}$$

Since the diffusion rate depends on r, the initial distribution over *r* can be perturbed during the transient process. However, the initial radial distribution must be reproduced in the steady-state.

For an initial condition given by a delta function at $r = r_0$, and $\theta = 0$, the solution for small *t* can be approximated by [39]

$$f(t, r_0, \theta) \approx \frac{r_0}{\sqrt{4\pi\kappa t}} \exp\left(\frac{r_0^2 \theta^2}{4\kappa t}\right) \tag{24}$$

This solution corresponds to an infinite domain, with boundary conditions $f \to 0$ as $\theta \to \pm\infty$. Thus, the solution is valid at small *t*, compared to $\kappa/r_0^2$, since for these times, the diffusion has not yet reached the point $\theta = \pi$. The singularity of the initial condition disappears, and the solution becomes smooth with respect to $\theta$, as seen in the exponential decay of the Fourier coefficients in (21). For large *t*, only a few terms in the sum are sufficient for a reasonable comparison with computational results.

We have solved the ring diffusion problem using tensor (18) in the Cartesian space for $x_0 = 2$ with a computational box of size -3<x,y<3 with Basilisk and COMSOL. For the selected values of $\delta \ll 1$, the boundary conditions at the box sides did not affect the solution.

Figure 13 shows an example of simulations using Basilisk implicit solver. Small negative values persist in the implicit code but can be reduced using limiters. The limiters help prevent the negative values of *f* that result from the off-diagonal diffusion coefficients, as suggested in Ref [40]. We have tested both implicit and explicit solvers in Basilisk C and have confirmed that the AMR solvers conserve the number of particles, just like the uniform grid solvers.

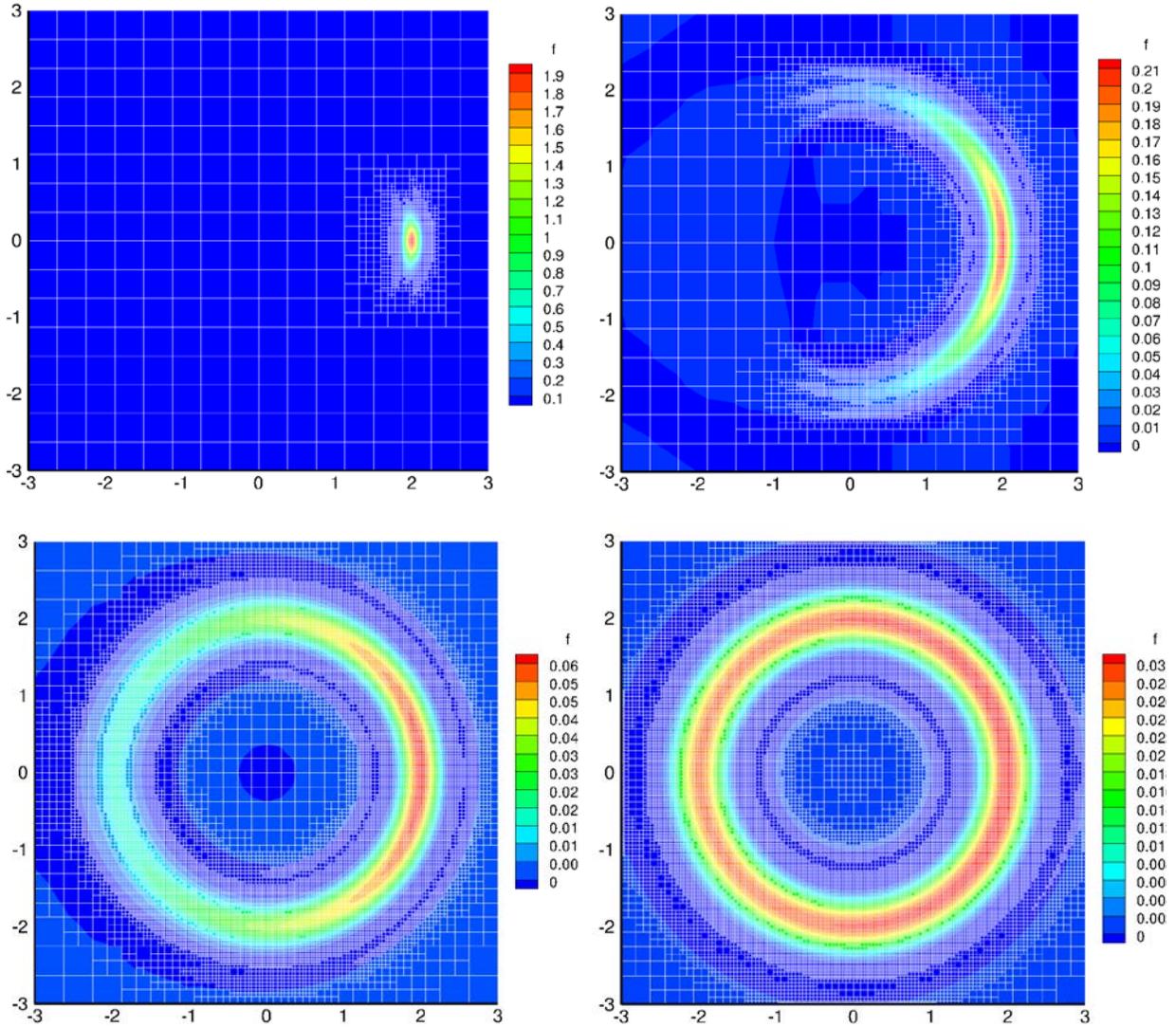

*Figure 13: Contours of f and adapted Cartesian mesh at dimensionless times 0, 2, 10, and 30 for the ring diffusion problem at $\delta = 0.1$, κ=0.5 obtained with Basilisk.*

Quantitative comparison of simulation results versus analytical solutions are shown in Figure 14 and Figure 15.

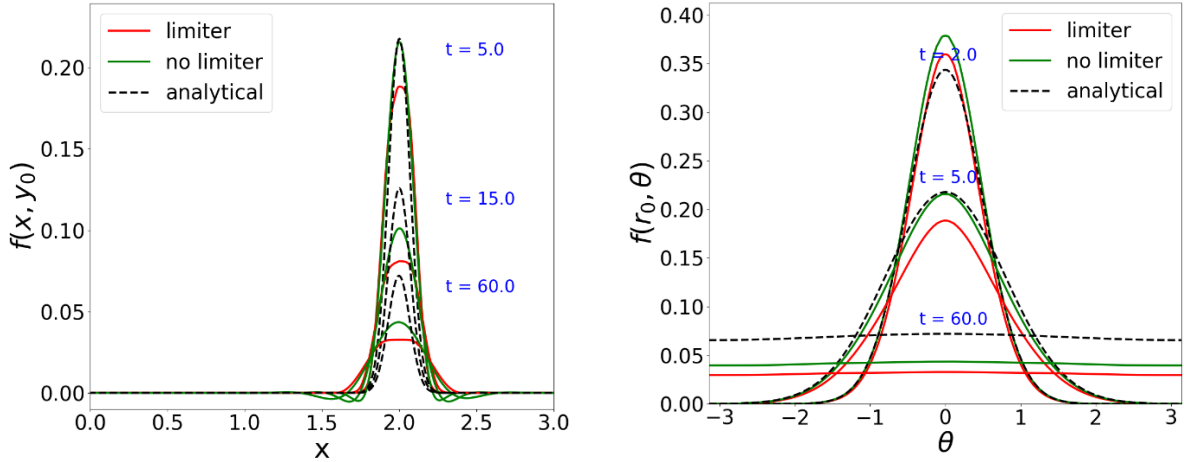

*Figure 14: Distributions along the radius (left) and along x for a narrow Gaussian $\delta = 0.1$*

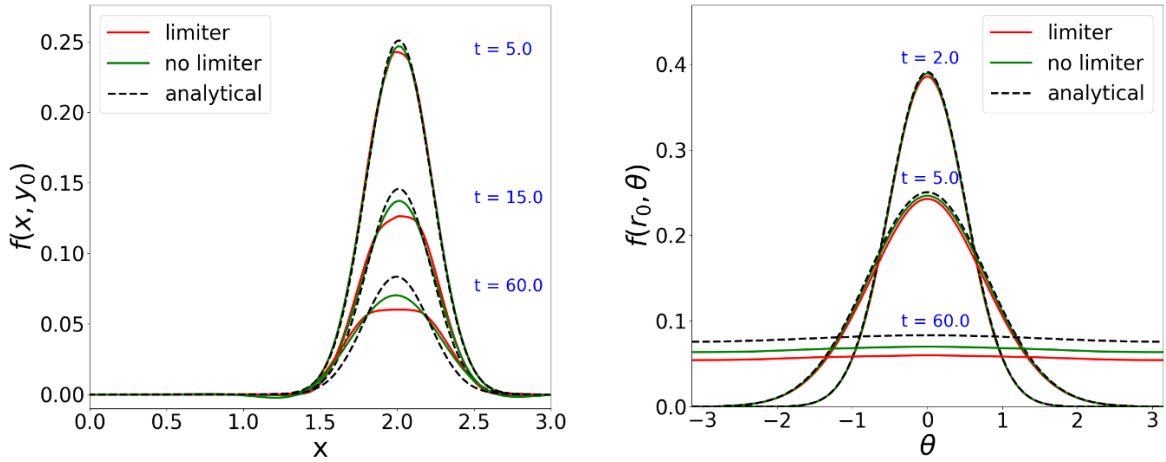

*Figure 15: Distributions along the radius (left) and along x for a broad Gaussian $\delta = 0.3$*

We found that limiters are effective at reducing the unphysical negative values in the distribution function but they come with trade-offs. Most limiters, like the generalized minmod limiter, create artificial diffusion, while others, like the superbee limiter, are anti-diffusive, both of which warp the distribution function at steady state. In comparison, the lack of a limiter leads to the appearance of significant negative values (~10% of the peak positive values) for very narrow initial distributions. For broader initial distribution, the non-limited numerical solution does not show large negative values, while more faithfully preserving the distribution shape, and thus might be preferable in most practical cases.

**Acknowledgments**

This work was supported by the NSF project OIA-1655280 and DOE project DE-SC0021391.

# References

[1] A. Garscadden, Ionization Waves in Glow Discharges, in *Gaseous Electronics* (1978)

[2] V.I. Kolobov, Striations in rare gas plasmas, J. Phys. D: Appl. Phys. **39** (2006) R487

[3] Y.-X. Liu, E. Schungel, I. Korolov, Z. Donko, Y.-N. Wang and J. Schulze, Experimental observation and computational analysis of striations in electronegative capacitively coupled radio-frequency plasmas, Phys. Rev. Lett. **116** (2016) 255002

[4] M. Tahiyat, J. Stephens, V. Kolobov and T. Farouk, Striations in moderate pressure dc driven nitrogen glow discharge, J. Phys. D: Appl. Phys. **55** (2022) 085201

[5] D. W. Swain and S. C. Brown, Moving Striations in a Low-Pressure Argon Discharge, Physics of Fluids **14**, 1383 (1971)

[6] K. Rohlena, T. Růžička & L. Pekárek, A theory of the low current ionization waves (striations) in inert gases, Czechoslovak Journal of Physics B **22**, 920 (1972)

[7] L.D. Tsendin, Ionization kinetics and ionization waves in neon, Soviet Physics - Technical Physics **27** (1982) 407, 412.

[8] V.A. Shveigert, Electron distribution function in inert gas in slightly modulated static electric field, Sov. J. Plasma Phys. **15** (1989) 714.

[9] Y. B. Golubovskii, V. I. Kolobov, and V. O. Nekuchaev, On electron bunching and stratification of glow discharges, Phys. Plasmas 20, 101602 (2013)

[10] N A Dyatko, I V Kochetov and V N Ochkin, Influence of the ionization process on characteristics of spatial relaxation of the average electron energy in inert gases in a uniform electric field, Phys. Rev. E **104** (2021) 065204

[11] D. Levko, Electron kinetics in standing and moving striations in argon gas, Phys. Plasmas **28**, 013506 (2021)

[12] R. R. Arslanbekov and V. I. Kolobov, Implicit and coupled fluid plasma solver with adaptive Cartesian mesh and its applications to non-equilibrium gas discharges, Plasma Sources Sci. Technol. **30** (2021) 045013.

[13] V. I Kolobov, R. R Arslanbekov, D. Levko and V. A. Godyak, Plasma stratification in radio-frequency discharges in argon gas, J. Phys. D: Appl. Phys. **53** (2020) 25LT01

[14] Yu. B. Golubovskii, V. A. Maiorov, V. O. Nekutchaev, J. Behnke, and J. F. Behnke, Kinetic model of ionization waves in a positive column at intermediate pressures in inert gases, Phys. Rev. E **63** (2001) 036409

[15] S. Arndt, F. Sigeneger, H. Testrich, and C. Brandt, Self-Consistent Analysis of the Spatial Relaxation of a Disturbed Neon Glow Discharge, Plasma Chemistry and Plasma Processing **25** (2005) 567

[16] A. V. Fedoseev and G. I. Sukhinin, A Self-Consistent Kinetic Model of the Stratification of Plane and Spherical Low-Pressure Discharges in Argon, Plasma Physics Reports **30** (2004) 1061.

[17] G. I. Sukhinin and A. V. Fedoseev, A Self-Consistent Kinetic Model of the Effect of Striation of Low-Pressure Discharges in Inert Gases, High Temperature. **44** (2006) 157.

[18] U. Kortshagen, C. Busch and L. D. Tsendin, On simplifying approaches to the solution of the Boltzmann equation in spatially inhomogeneous plasmas, Plasma Sources Science and Technology **5** (1996) 1-17

[19] G. Giorgiani, H. Bufferand, F. Schwander, E. Serre, P. Tamain, A high-order non field-aligned approach for the discretization of strongly anisotropic diffusion operators in magnetic fusion, Computer Physics Communications **254** (2020) 107375

[20] L. D. Tsendin, Nonlocal electron kinetics in gas-discharge plasma, Phys.-Usp. **53** (2010) 133.

[21] V. I. Kolobov, Advances in electron kinetics and theory of gas discharges, Physics of Plasmas **20**, 101610 (2013)

[22] V.I. Kolobov; R.R. Arslanbekov, Simulation of electron kinetics in gas discharges, IEEE Transactions on Plasma Science **34** (2006) 895

[23] Steven Rosenberg, *The Laplacian on a Riemannian manifold: an introduction to analysis on manifolds*, Cambridge University Press (1997)

[24] R. E. Robson, R. D. White, and M. Hildebrandt, One hundred years of the Franck-Hertz experiment, Eur. Phys. J. D **68** (2014) 188; DOI: 10.1140/epjd/e2014-50342-9

[25] N A Dyatko, I V Kochetov, and V N Ochkin, Peculiarities of spatial relaxation of the mean electron energy in inert gases and their mixtures in a uniform electric field, Plasma Sources Sci. Technol. **29** (2020) 125007

[26] COMSOL Multiphysics Software, Available: https://www.comsol.com/.

[27] "Basilisk C," [Online]. Available: http://basilisk.fr/.

[28] C. Yuan, E. A. Bogdanov, S. I. Eliseev, and A. A. Kudryavtsev, 1D kinetic simulations of a short glow discharge in helium, Physics of Plasmas **24**, 073507 (2017)

[29] Yan Chai, Jingfeng Yao, E A Bogdanov, A A Kudryavtsev, Chengxun Yuan and Zhongxiang Zhou, Formation of inverse EDF in glow discharges with an inhomogeneous electric field, Plasma Sources Sci. Technol. **30** (2021) 095006

[30] V. A. Godyak, B. M. Alexandrovich, and V. I. Kolobov, Measurement of the electron energy distribution in moving striations at low gas pressures, Phys. Plasmas **26**, 033504 (2019)

[31] Yu. B. Golubovskii, V. S. Nekuchaev, and N. S. Ponomarev, Trapped and free electrons in the near-anode region of a striated discharge, Techical Physics **43** (1998) 288

[32] L.D. Tsendin, Electron distribution function in a weakly ionized plasma in an inhomogeneous electric field. II - Strong fields (energy balance determined by inelastic collisions), Soviet Journal of Plasma Physics **8** (1982) 228.

[33] A W Cooper, J R M Coulter and K G Emeleus, Simultaneous Occurrence of Moving and Standing Waves in a Positive Column, Nature **181** (1958) 1326

[34] V. I. Kolobov and L.D. Tsendin, Analytic model of the cathode region of a short glow discharge in light gases, Phys. Rev. A **46** (1992) 7837

[35] Yu. Golubovskii, T. Gurkova, S. Valin, Resonant behavior of the electron component of the plasma and stratification of the positive column of a gas discharge, Plasma Sources Sci. Technol. **30** (2021) 115001

[36] T. Yang and Y. Wang, A new tailored finite point method for strongly anisotropic diffusion equation on misaligned grids, Applied Mathematics and Computation **355** (2019) 85

[37] P. Sharma and G. W. Hammett, A fast semi-implicit method for anisotropic diffusion, J. Comput. Phys. **230** (2011) 4899

[38] N. Crouseilles, M. Kuhn, and G. Latu, Comparison of Numerical Solvers for Anisotropic Diffusion Equations Arising in Plasma Physics, J. Sci. Comput. **65** (2015) 1091

[39] S. Cordier, B. Lucquin-Desreux, and S. Mancini, Focalization: A Numerical Test for Smoothing Effects of Collision Operators, Journal of Scientific Computing, **24** (2005) 311

[40] P. Sharma and G. W. Hammett, Preserving monotonicity in anisotropic diffusion, J. Comput. Phys. **227** (2007) 12.